\documentclass[twocolumn,aps,prb,widetext,showpacs,superscriptaddress]{revtex4-1}
\usepackage{mathrsfs,bm,multirow,amsmath,amsfonts,amssymb,array,booktabs,float}
\usepackage{graphicx,times,graphics,color,epsfig}
\begin{document}
\title{Composition dependent band offsets of ZnO and its ternary alloys}
\author{Haitao Yin}
\affiliation{Key Laboratory for Photonic and Electronic Bandgap Materials of Ministry of Education, School of Physics and Electronic Engineering, Harbin Normal University, Harbin 150025, China}
\affiliation{Department of Physics and the Center of Theoretical and Computational Physics, The University of Hong Kong, Pokfulam Road, Hong Kong SAR, China}
\author{Junli Chen}
\affiliation{Key Laboratory for Photonic and Electronic Bandgap Materials of Ministry of Education, School of Physics and Electronic Engineering, Harbin Normal University, Harbin 150025, China}
\author{Yin Wang}
\email{yinwang@hku.hk}
\affiliation{Department of Physics and the Center of Theoretical and Computational Physics, The University of Hong Kong, Pokfulam Road, Hong Kong SAR, China}
\affiliation{The University of Hong Kong Shenzhen Institute of Research and Innovation, Shenzhen, Guangdong 518057, China}
\author{Jian Wang}
\affiliation{Department of Physics and the Center of Theoretical and Computational Physics, The University of Hong Kong, Pokfulam Road, Hong Kong SAR, China}
\affiliation{The University of Hong Kong Shenzhen Institute of Research and Innovation, Shenzhen, Guangdong 518057, China}
\author{Hong Guo}
\affiliation{Center for the Physics of Materials and Department of Physics, McGill University, Montreal, Quebec  H3A 2T8, Canada}
\affiliation{Department of Physics and the Center of Theoretical and Computational Physics, The University of Hong Kong, Pokfulam Road, Hong Kong SAR, China}

\date{\today}

\begin{abstract}
We report the calculated fundamental band gaps of \emph{wurtzite} ternary alloys Zn$_{1-x}$M$_x$O (M=Mg, Cd) and the band offsets of the ZnO/Zn$_{1-x}$M$_x$O heterojunctions, these II-VI materials are important for electronics and optoelectronics. Our calculation is based on density functional theory within the linear muffin-tin orbital (LMTO) approach where the modified Becke-Johnson (MBJ) semi-local exchange is used to accurately produce the band gaps, and the coherent potential approximation (CPA) is applied to deal with configurational average for the ternary alloys. The combined LMTO-MBJ-CPA approach allows one to simultaneously determine both the conduction band and valence band offsets of the heterojunctions. The calculated band gap data of the ZnO alloys scale as $E_g=3.35+2.33x$ and $E_g=3.36-2.33x+1.77x^2$ for Zn$_{1-x}$Mg$_x$O and Zn$_{1-x}$Cd$_x$O, respectively, where $x$ being the impurity concentration. These scaling as well as the composition dependent band offsets are quantitatively compared to the available experimental data. The capability of predicting the band parameters and band alignments of ZnO and its ternary alloys with the LMTO-CPA-MBJ approach indicate the promising application of this method in the design of emerging electronics and optoelectronics.
\end{abstract}

\maketitle

\emph{NOTE:} This article has been accepted by Scientific Reports in a revised form (www.nature.com/articles/srep41567).

\section{Introduction}
Zinc oxide (ZnO), having a direct band gap ($E_g$) of 3.37 eV and a high exciton binding energy of about 60 meV at room temperature,\cite{Reynolds, Huang} is a promising material for electronics and optoelectronics, including applications to solar cells,\cite{Nuruddin} light-emitting diodes\cite{Kong,Nakahara} and ultraviolet micro-lasers.\cite{Tang} In electronics technology, semiconductor heterojunctions are commonly used to provide potential barriers and/or quantum wells that tune and control carrier transport. Composition controlling techniques are the most widely applied method to alter the band gap of materials in the so called ``band engineering" to build heterojunctions. For ZnO, composition doping of Mg and Cd atoms has been used to increase or decrease the band gap, respectively.\cite{Ohtomo, Bendersky, Zhang, Su}

It could be quite tedious and difficult to experimentally measure the band gaps of semiconductor alloys and band offsets of their heterojunctions over the entire doping concentration, therefore first principles theoretical calculations and predictions are of great fundamental interest as well as practical relevance. In particular, as nanoelectronic devices are reaching the sub-10 nm scale, atomistic first principles prediction of band parameters of semiconductor materials and heterojunctions is becoming very important in order to design and/or select new materials that have desired properties. However, it has been a serious challenge for existing first principles methods to accurately predict band information of semiconductor alloys and heterojunctions for several reasons. First of all, let's consider calculating the band offset of a heterojunction A$_{1-x}$B$_x$/A where A is a semiconductor and $x$ the doping concentration of some impurity specie B. If $x$ is small as is usually the case, e.g. $x=0.1\%$, one has to calculate a system containing at least $1000$ host atoms A in order to accommodate just a single impurity atom B. Namely the total number of atoms are very large when $x$ is small which makes first principles analysis extremely difficult to do. Second, one has to perform a configurational average of the calculated results because the doped impurity atoms are randomly distributed in the sample, and the configurational average is much more computationally costly. Finally, it is well-known that the first principles methods of density functional theory (DFT)\cite{DFT1, DFT2} with local-density approximation (LDA)\cite{LDA, LDACA, Vosko80} and generalized gradient approximation (GGA)\cite{Perdew92b, GGAPBE96} underestimate the band gaps of semiconductors and, without a correct band gap, the accuracy of the calculated band offset could be questionable. For these reasons, the composition dependent band offsets of many important semiconductors, including that of ZnO and its ternary alloys, have not been theoretically investigated to a satisfactory level.

During the past decades, considerable theoretical efforts have been devoted to solve the above band gap and configurational average issues. Among them the combined DFT approach of coherent potential approximation (CPA)\cite{CPA} and modified Becke-Johnson (MBJ) semilocal exchange\cite{MBJ} under the linear muffin-tin orbital (LMTO) scheme,\cite{LMTO, LMTOBOOK} has been successful in calculating the band information of many semiconductors, their alloy and heterojunctions. So far, the LMTO-CPA-MBJ approach has been applied to predict the band gaps of all group III-V semiconductors,\cite{Wang1} the alloy InGaN,\cite{Cesar} AlGaAs,\cite{Wang2} the group IV semiconductor alloys SiGeSn,\cite{Zhuzhen} as well as the band offset of group III-V semiconductor heterojunctions GaAs/AlGaAs.\cite{Wang2} Quantitative comparison between the calculated results and measured data were very satisfactory. However, a more challenging and important semiconductor family - \emph{wurtzite} group II-VI semiconductors, has not been investigated by the LMTO-CPA-MBJ approach yet.

In this work, we have calculated the band gap of \emph{wurtzite} group II-VI semiconductor alloys Zn$_{1-x}$M$_x$O (where M = Mg or Cd) and the band offset of ZnO/Zn$_{1-x}$M$_x$O using the LMTO-CPA-MBJ approach. The calculated band gaps of Zn$_{1-x}$Mg$_x$O and both conduction and valence band offsets of ZnO/Zn$_{1-x}$Mg$_x$O heterojunction exhibit a linear function of Mg concentration, while a nonlinear behavior of the band gaps of ZnCdO and band offsets of ZnO/ZnCdO is predicted. The calculated band gaps and band offset of heterojunctions agree well with the corresponding experimental measurements. Our calculated band gap data of ZnO alloys are well fit to the following scaling expressions: $E_g=3.35+2.33x$ for alloy Zn$_{1-x}$Mg$_x$O, and $E_g=3.36-2.33x+1.77x^2$ for alloy Zn$_{1-x}$Cd$_x$O. Together with the previously reported results on group III-V semiconductors\cite{Cesar, Wang1, Wang2} and group IV semiconductors,\cite{Zhuzhen} we can confidently expect that the band parameters and band alignments of most conventional semiconductors can be well predicted from first principles by LMTO-CPA-MBJ approach.

\section{Results and Discussions}
\begin{figure}
\includegraphics[width=\columnwidth]{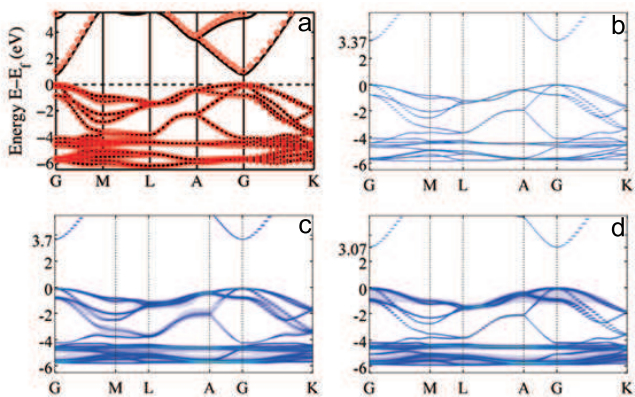}\\
\caption{(Color online) (a) The calculated LDA band structure of pure ZnO. Black solid line was obtained by the planewave electronic package \texttt{VASP}, and red circles by \texttt{NANODSIM}. (b) The calculated MBJ band structure of pure ZnO crystal, giving the experimental band gap of 3.37 eV. (c) The calculated CPA band structure of the Zn$_{0.85}$Mg$_{0.15}$O alloy, the band gap is increased to 3.7 eV. (d) The calculated CPA band structure of the Zn$_{0.85}$Cd$_{0.15}$O alloy, the band gap is decreased to 3.07 eV compared to that of pure ZnO.}
\label{fig1}
\end{figure}

\textbf{Band structure of pure semiconductor ZnO.}
To carry out DFT calculations within the LMTO scheme, atomic sphere approximation (ASA) should be employed for space filling with placing atom and vacancy spheres at appropriate locations.\cite{LMTO, LMTOBOOK} Different ASA scheme, i.e., different sphere positions and radii, will affect the calculated electronic structure. To verify our ASA scheme of the LMTO method, we calculated the band structure of pure ZnO crystal at the level of local LDA to compare with that obtained by the Vienna Ab initio Simulation Package (\texttt{VASP}).\cite{VASP1, VASP2} As shown in Fig.~\ref{fig1}(a), the agreement is excellent - suggesting the appropriateness of our ASA scheme. Note that the calculated LDA band gap, though consistent with that reported in the literature,\cite{Schroer,Janotti} is underestimated which is a well-known issue of LDA.

Having obtained the correct band structure of pure ZnO by LMTO at the LDA level, we further determine the $c$-values of the MBJ potential (See Methods Section) in order to produce the experimental $E_g$, 3.37 eV, of pure ZnO. Fig.~\ref{fig1} (b) shows the MBJ result, and the opening of the $E_g$ as campared to our LDA result as well as the previously reported LDA $E_g$ of 0.74 eV\cite{Schroer, Janotti} and GGA $E_g$ of 0.80 eV,\cite{Oba} is evident. Till now, we have decided all the parameters for the LMTO-CPA-MBJ approach to produce the experimental band gap of pure ZnO, including the LMTO related ASA parameters for sphere positions and radii and the MBJ related parameter $c$-values for increasing the band gap of semiconductors. More complicated ASA scheme and more $c$-values for different spheres may produce much more accurate band parameters, we would note that our choice is a good compromise between simplicity and reasonable accuracy. Though a few parameters need to be pre-determined in our first principle calculations, it is acceptable since we are dealing a very challenging problem and these parameters can be easily obtained. In the further calculations of the band gap of semiconductor alloys and band offset of semiconductor heterojunctions, we will fix all the parameters obtained in the pure ZnO calculations.

\textbf{Band gaps of ZnO ternary alloys.}
Having correctly determined the band gap of pure ZnO crystal, exactly the same ASA scheme and the MBJ semi-local exchange potential were used to calculate the electronic structures of ZnO alloys. Note that when ZnO crystal is doped with (substitutional) impurity atoms at random sites, translational symmetry is broken and momentum \textbf{k} is no-longer a good quantum number such that Bloch's theorem no longer exists. In our CPA calculations,\cite{CPA, Eric} the spirit of the approach is to construct an effective medium by completing configurational average over the random disorder that restores the translational invariance, as such one can again speak of a ``band structure" with bands that are broadened by impurity scattering.

Fig.~\ref{fig1}(c,d) plots the calculated CPA band structure of ZnO alloy with 15\% MgO and 15\% CdO composition, respectively. From Fig.\ref{fig1}(c), we obtain $E_g=3.70$ eV that is $0.33$ eV higher than that of pure ZnO. This is very close to the experimentally measured value of $E_g=3.68$ eV for Zn$_{0.85}$Mg$_{0.15}$O.\cite{Su} For Zn$_{0.85}$Cd$_{0.15}$O, the calculated band gap is $E_g=3.07$ eV, namely, $0.3$ eV smaller than that of pure ZnO.

\begin{figure}
\includegraphics[width=\columnwidth]{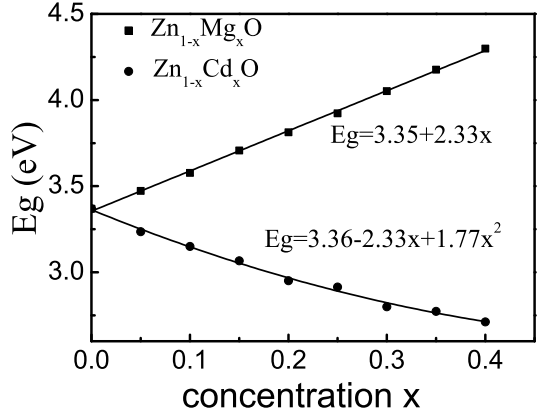}\\
\caption{Solid squares and dots are calculated band gap $E_g$ of the Zn$_{1-x}$Mg$_x$O and Zn$_{1-x}$Cd$_x$O ternary alloys, respectively. The increase of Mg (Cd) composition enlarges (narrows) the band gap of these \emph{wurtzite} alloys with the Mg (Cd) composition from 0 to 40\%.}
\label{fig2}
\end{figure}

We have calculated $E_g$ of wurtzite ZnO alloys versus impurity concentration $x$ up to $x=40\%$, for Zn$_{1-x}$M$_x$O where M = Mg or Cd, as shown in Fig.~\ref{fig2}. Beyond $x=40\%$, it is known experimentally\cite{Ohtomo, Bendersky} that ZnO alloy tends to form a \emph{cubic} phase (which is not the focus of this work). For Zn$_{1-x}$Mg$_x$O alloys, our calculated $E_g$ increases linearly with the MgO concentration as $E_g=3.35+2.33x$ (the solid squares line in Fig.\ref{fig2}). This scaling is consistent with the experimental observations of $E_g=3.4+2.3x$ in Ref.~\onlinecite{Adachi}, and $E_g=3.37+2.51x$ in Ref.~\onlinecite{Koike}. According to our calculations, $E_g$ of the \emph{wurtzite} Zn$_{1-x}$Mg$_x$O alloy can be tuned to 4.29 eV at $x=40\%$. Our fitted linear slope agrees outstandingly with previous calculated value of 2.03 by LDA,\cite{Franz} though LDA gives underestimated band gaps. Moreover, if quadratic fitting is applied to our calculated data, the coefficient of the $x^2$ term gives the bowing parameter of 0.18 eV, which is close to the previous calculated bowing parameter of 0.44.\cite{Schleife}
For the Zn$_{1-x}$Cd$_x$O alloys, we found that $E_g$ changes with the Cd concentration as $E_g=3.36-2.33x+1.77x^2$ which is nonlinear (the dots line line in Fig.\ref{fig2}), and the $E_g$ of the \emph{wurtzite} Zn$_{1-x}$Cd$_x$O alloy can be narrowed to 2.71 eV when $x=40\%$. Here, the coefficient of the $x^2$ term gives the bowing parameter of 1.77 eV, which agrees extremely well with the experimental value of 1.75eV.\cite{Ishihara} Note that previous calculated bowing parameter was considerably smaller, 1.21 eV in Ref.~\onlinecite{Zhang2} and 1.03 eV in Ref.~\onlinecite{Fan}.

\begin{figure}
\includegraphics[width=\columnwidth]{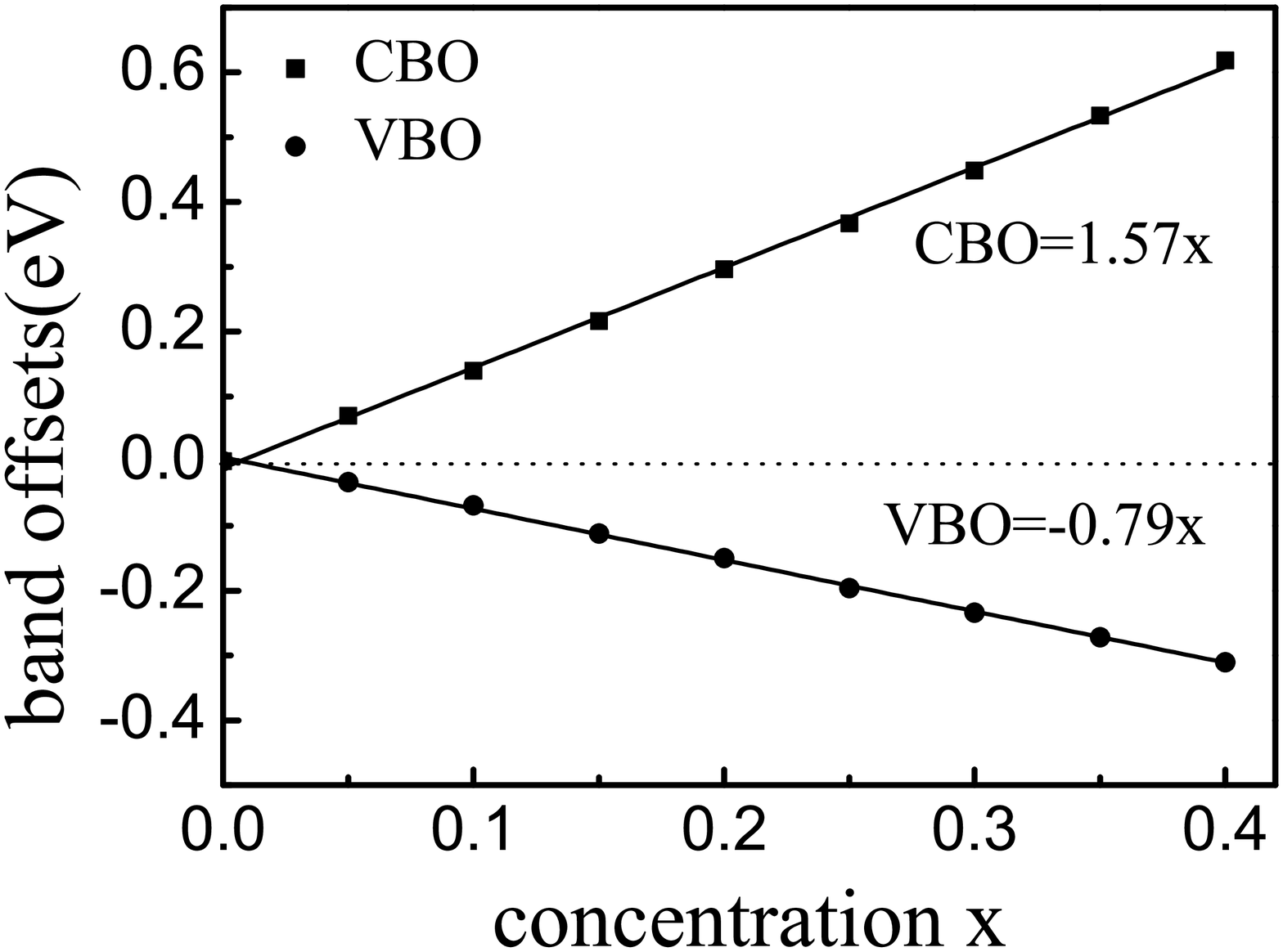}\\
\caption{Solid squares and dots present the CBO and VBO of the ZnO/Zn$_{1-x}$Mg$_x$O heterojunctions for $x<40\%$, respectively.}
\label{fig3}
\end{figure}

\textbf{Band offsets of ZnO and its ternary alloys.}
Having obtained accurate $E_g$ for the ZnO alloys, a very important task is to predict the band offsets for the ZnO/Zn$_{1-x}$M$_x$O heterojunctions. Since our method can accurately predict the band gaps of pure ZnO and its ternary alloys, both the conduction band offset (CBO) and the valence band offset (VBO) can be calculated in one shot. In comparison, for those calculations where $E_g$ was not correctly obtained (such as using LDA/GGA functional), the conduction band offset could not be predicted unless one amends $E_g$ after the valence band offset is calculated. Fig.~\ref{fig3} plots the calculated band offsets of ZnO/Zn$_{1-x}$Mg$_x$O heterojunctions for $x<40\%$. The band offsets linearly scale as VBO$=-0.79x$ eV and CBO$=+1.57x$ eV. Here the minus sign in VBO means that the valence band maximum of the ZnMgO alloy is lower than that of the pure ZnO; while the plus sign in CBO indicates the conduction band minimum of the ZnMgO is higher than that of the pure ZnO. From the fitted curves (black lines), the band offset ratio (VBO:CBO) is about 2. This is in the range of the experimentally reported values:  3:2 to 7:3 as reported in Ref.~\onlinecite{Coli}, and 1.5 to 2 as reported in Ref.~\onlinecite{Zhang}. At Mg concentration of 15\%, our calculated CBO (VBO) is 0.22 eV (0.11 eV), which agrees with the experimentally measured data of  0.16 eV (0.09 eV).\cite{Zhang}

\begin{figure}
\includegraphics[width=\columnwidth]{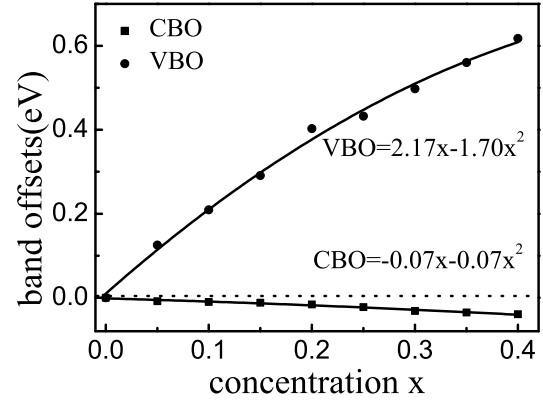}\\
\caption{Solid squares and dots present the CBO and VBO of the ZnO/Zn$_{1-x}$Cd$_x$O heterojunctions for $x<40\%$, respectively.}
\label{fig4}
\end{figure}

Fig.~\ref{fig4} shows the calculated band offsets of ZnO/Zn$_{1-x}$Cd$_x$O for $x<40\%$, and a nonlinear relationship with the Cd concentration is obtained: VBO$=2.17x-1.70x^2$ eV and CBO$=-0.07x-0.07x^2$ eV. The negative coefficient of $x$ for CBO indicates that the conduction band minimum of the ZnCdO is lower than that of the pure ZnO, while the positive coefficient of $x$ for VBO shows a higher valence band maximum of the ZnCdO. In particular, our calculated VBO values of 0.20 eV at $x=10\%$ and 0.104 eV at $x=5\%$ agree very well with the experimental values of 0.203 eV at $x=9.6\%$\cite{Yao} and $0.17\pm 0.03$ eV at $x=5\%$\cite{Chenexp}.

\begin{figure}
\includegraphics[width=\columnwidth]{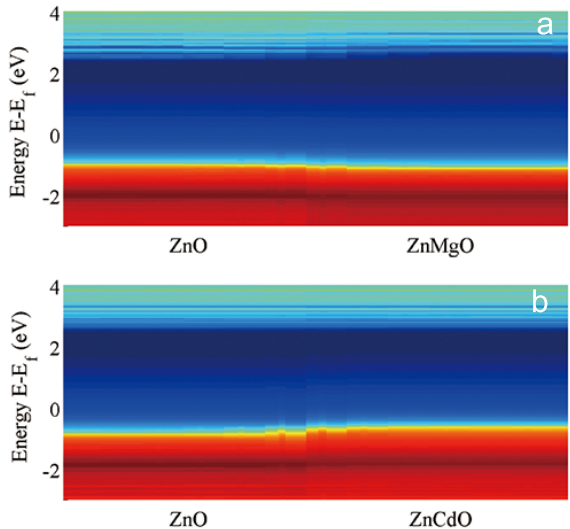}\\
\caption{(Color online) The local projected DOS of: (a) ZnO/Zn$_{0.85}$Mg$_{0.15}$O and (b) ZnO/Zn$_{0.85}$Cd$_{0.15}$O heterojunctions. The cooler color indicates lower DOS. The blue (darker) region intuitively shows the band gap of the semiconductors. ZnO and its ternary alloys ZnMgO and ZnCdO both forms type-I band alignment.}
\label{fig5}
\end{figure}
Fig.~\ref{fig5} intuitively show the band alignment of ZnO/Zn$_{0.85}$M$_{0.15}$O (M=Mg, Cd) by plotting the local projected DOS\cite{LPDOS1, LPDOS2} along the heterojunction direction. It can be clearly observed that the pure ZnO crystal formed type-I band alignment heterojunction with its Mg and Cd doped ternary alloys, and the band gap of pure ZnO is smaller than that of ZnMgO but larger than that of ZnCdO.

\textbf{Discussions.}
Based on our LMTO-CPA-MBJ calculations on ZnO and its alloys, a general procedure for predicting band gaps of semiconductor alloys and band offsets of heterojunctions is in order. (1) Choose a proper ASA scheme and determine the $c$-value of the MBJ semi-local exchange to obtain accurate band structures of the pure semiconductor; (2) calculate the band information of semiconductors alloys by CPA; (3) calculate the band offset of the semiconductor heterojunction with corrected determined band gaps, from which the valence and conduction band offset can be predicted in one shot. Since a major problem of semiconductor technology is to design materials having desired electronic structure, the theoretical procedure summarized here should be applicable to a wide range of research issues.

The combined LMTO-CPA-MBJ approach is firstly applied on the band gap calculation of group III-V semiconductor alloy In$_x$Ga$_{1-x}$N, the results are in excellent agreement with the measured data for the entire range of $0\leq x \leq 1$.\cite{Cesar} Later on, the approach is implemented in \texttt{NANODSIM} software package.\cite{Eric} Using \texttt{NANODSIM} Ref.~\onlinecite{Wang1} calculated the band structures and effective masses of all the zinc-blende group III-V semiconductors with LMTO-MBJ approach, and very good quantitative comparison with the experimental data is made; Ref.~\onlinecite{Wang2} for the first time calculated the band offset of GaAs/Al$_x$Ga$_{1-x}$As heterojunctions for the entire range of the Al doping concentration $0< x \leq 1$ using LMTO-CPA-MBJ approach, both the conduction band offset (CBO) and the valence band offset (VBO) agree very well with many experiments; Ref.~\onlinecite{Zhuzhen} calculated the composition dependent band gaps of group IV ternary alloys SiGeSn, in which three elements locate on the same site with certain probabilities and their average physical quantities were dealt with CPA. These previously reported results show that the LMTO-CPA-MBJ approach can accurately calculate the band information of group III-V and group IV semiconductors and their binary and ternary alloys and heterojunctions. In this work, we further applied this approach on a more challenging system - \emph{wurtzite} group II-VI semiconductors, and the obtained results are quantitatively compared with the experimental measurements.

We used the effective medium CPA method under LMTO scheme to calculate the electronic structures of semiconductors and alloys, which does not capture any atomic relaxations. Considering that we are dealing a very challenging problem, which needs to solve the doping, configurational average, band gap underestimation and computational cost issues in one calculation, to this end, the combined LMTO-CPA-MBJ approach is a reasonably compromised method. Moreover, for many semiconductor alloys, their compositions have very close lattice parameters, making the structural change negligible. It could be expected that this approach can be used in the prediction of band parameters of wide range of semiconductors.

\section{Summary}
Using a state-of-the-art atomistic approach, we have calculated the band gap of \emph{wurtzite} ZnO based group II-VI ternary alloys with Mg or Cd compositions, and the band offsets of ZnO and its ternary alloy heterojunctions. The calculated band gaps of the ZnO alloys scale as $Eg=3.35+2.33x$ and $Eg=3.36-2.33x+1.77x^2$ for Zn$_{1-x}$Mg$_x$O and Zn$_{1-x}$Cd$_x$O, respectively. Further calculations on the band offsets show that the pure ZnO formed type-I band alignment heterojunction with its Mg and Cd doped ternary alloys. The calculated band gaps of the alloys and band offset of the heterojunctions quantitatively agree with the experimental measurements for $x<40\%$ where the ZnO and its Mg and Cd doped alloys are in  the \emph{wurtzite} phase. The success of the LMTO-CPA-MBJ on predicting the band gaps and band offsets of group II-VI semiconductor alloys and heterojunctions, together with its capability on calculating and predicting the band gaps and band offsets of group IV\cite{Zhuzhen} and group III-V\cite{Wang1, Wang2, Cesar} semiconductors, make us can confidently conclude that the LMTO-CPA-MBJ approach can be widely used to predict the semiconductor band parameters and band alignments from first principles and will be very useful to the design of emerging electronics and optoelectronics.

\section{Methods}

Our calculations are based on the LMTO-CPA-MBJ self-consistent approach where DFT is carried out in the LMTO scheme with the atomic sphere approximation (ASA),\cite{LMTO, LMTOBOOK} as implemented in the \texttt{NANODSIM} software package.\cite{Eric} For ASA, vacancy spheres were placed at appropriate locations for space filling [See TABLE~\ref{tab1} for details].\cite{Cesar} In particular, the simulation cell is filled by slightly overlapped spheres within ASA, and the cell volume of the whole simulated system $V_{cell}$ should be equal to the total volume of the atom and vacancy spheres $V_{spheres}$. With the sphere position of the \emph{wurtzite} structure in TABLE~\ref{tab1}, we optimized the sphere radii by reproducing the electronic band structures of pure ZnO crystals. Afterward, to deal with the impurity doped ZnO alloy, the statistical effective medium CPA theory\cite{CPA} is applied which allows us to obtain the configurational averaged results without individually computing each atomic configuration.

We use the MBJ semi-local exchange that can accurately determine the band gap $E_g$ of the semiconductors.\cite{MBJ} Following the original paper,\cite{MBJ} the MBJ semi-local exchange potential has the following form,
\begin{equation}\label{eq-MBJ}
v_{x,\sigma}^{MBJ}(r)=cv_{x,\sigma}^{BR}(r)+(3c-2)\frac{1}{\pi}\sqrt{\frac{5}{12}}
\sqrt{\frac{2t_{\sigma}(r)}{\rho_{\sigma} (r)}},
\end{equation}
where subscript $\sigma$ is spin index, $\rho_{\sigma}$ is the electron density for spin channel $\sigma$. The quantity $t_{\sigma}$ is the kinetic energy density and $v_{x,\sigma}^{BR}(r)$ is the Becke-Roussel potential used in Ref.~\onlinecite{BR}. The above MBJ potential has two terms whose relative weight is given by a parameter $c$, which depends linearly on the square root of the average of $|\nabla \rho|/ \rho$. According to many previous studies,\cite{MBJ, Singh, spMBJ, Wang1, Wang2, Zhuzhen, Cesar} the calculated $E_g$ increases monotonically with $c$. For simplicity, we only used different $c$-values for real atom spheres and vacancy spheres, respectively. While the $c$-values can be self-consistently calculated using the electron density, in our calculations we fixed it to be $c=1.75$ and $c=1.13$ for vacancy spheres and real atom spheres respectively, which gave an $E_g$ of ZnO in excellent agreement with the experimental value (3.37 eV). Having obtained the correct band gap of ZnO, the same values of the $c$-parameters were used in all the subsequent calculations of the alloys and heterojunctions.

\begin{table}[t]
\caption{Positions of atomic spheres in the \emph{wurtzite} structure. $V_T$ and $V_O$ denote the vacancy spheres at the  tetrahedral center and octahedral center of the ZnO primitive cell, respectively. The optimized sphere radii of 1.1481~$\AA$, 0.7654~$\AA$, and 1.3023~$\AA$ are used for the real atom spheres, vacancies at the  tetrahedral center and vacancies at the octahedral center, respectively.}
\centering
\begin{tabular} {cccccc}
\hline\hline
Site    &Zn             		&O  							&$V_T$    						 &$V_O$  					\\ \hline
x        &$0$             		&$0$       						&$0$								 &$0$						\\
y        &$0$             		&$0$        						&$0$								 &$\sqrt{3}/3$				\\
z        &$0$             		&$u\cdot c$              			 &$\frac{1+u}{2}\cdot c$			&$u\cdot c/2$				\\
x        &$1/2$            		&$1/2$        					&$1/2$							 &$0$						\\
y        &$\sqrt{3}/6$   		&$\sqrt{3}/6$          			&$\sqrt{3}/6$						 &$\sqrt{3}/3$				\\
z        &$c/2$            		&$(\frac{1}{2}+u)\cdot c$     	&$u\cdot c/2$   					 &$\frac{1+u}{2}\cdot c$	\\
\hline\hline
\end{tabular}
\label{tab1}
\end{table}

The experimental lattice constants\cite{lattice, Jaffe} of hexagonal \emph{wurtzite} ZnO, $a=3.2495~\AA$,\cite{lattice} $c/a=1.602$,\cite{lattice} and $u=0.382c$,\cite{Jaffe} were used in our calculations for pure ZnO and all the alloys. The primitive cell of the \emph{wurtzite} structure was used to calculate the band structures and $E_g$ of ZnO and its alloys. To determine the band offset, (1120) supercell systems containing 8 unit cell of pure ZnO and 8 unit cell layers of ZnO alloy were used to calculate the potential profile along the heterojunction. A $10\times10\times10$ $k$-mesh and a $6\times6\times3$ $k$-mesh were used to sample the Brillouin Zone (BZ) for the primitive cells and the heterojunctions, respectively. The band gap values, as well as the energy position of the valence band maximum and conduction band minimum are read from the CPA band structure. The CPA band structure can intuitively trace the shape of the band structure with broadening due to impurity scattering, and provide more accurately and rich band information for the semiconductor and its alloys. The band offset values were calculated based on the electrostatistic potential profile of the heterojunction and the primitive cell, as more details in Ref.~\onlinecite{Wang2}.

\section*{Acknowledgements}
Y.W. is grateful to Drs. Yu Zhu, Lei Liu and Dongping Liu for useful discussions regarding the use of the \texttt{NANODSIM} software package. This work is supported by the University Grant Council (Contract No. AoE/P-04/08) of the Government of HKSAR (YW, JW, HG), the Open Project Program of Key Laboratory for Photonic and Electric Bandgap Materials (HY, YW), Special Program for Applied Research on Super Computation of the NSFC-Guangdong Joint Fund (YW, JW), Seed Funding Programme for Basic Research of HKU (YW) and NSFC (No. 11404273).

\bibliography{ref}

\end{document}